# Detectability of Tensor Perturbations Through CBR Anisotropy

Lloyd Knox[1,2] and Michael S. Turner[1,2,3]

[1]*NASA/Fermilab Astrophysics Center*
*Fermi National Accelerator Laboratory, Batavia, IL 60510-0500*

[2]*Department of Physics*
*The University of Chicago, Chicago, IL 60637-1433*

[3]*Department of Astronomy & Astrophysics*
*Enrico Fermi Institute, The University of Chicago, Chicago, IL 60637-1433*

**ABSTRACT**

Detection of the tensor perturbations predicted in inflationary models is important for testing inflation as well as for reconstructing the inflationary potential. We show that because of cosmic variance the tensor contribution to the square of the CBR quadrupole anisotropy must be greater than about 20% of the scalar contribution to ensure a statistically significant detection of tensor perturbations. This sensitivity could be achieved by full-sky measurements on angular scales of 3° and 0.5°.

Inflation has had almost as much impact on cosmology as the big-bang model itself. It provides an explanation for the flatness and smoothness of the observed Universe, as well as the inhomogeneity needed to seed the structure seen today. However, inflation has yet to be tested in a significant way. The keys to doing so are its three robust predictions: spatially-flat Universe (total energy density, including matter, radiation and vacuum energy equal to the critical density) and nearly scale-invariant spectra of scalar (density) [1] and tensor (gravity-wave) [2] metric perturbations. There is some evidence that the density of matter is close to unity [3] and several large-scale experiments are underway to search for the nonbaryonic dark matter that must exist if $\Omega_{\rm matter} = 1$. Scalar perturbations seed the formation of structure, and hence, observational data concerning the distribution of matter in the Universe can provide information about them. Today, tensor perturbations correspond to a stochastic background of gravitational waves that could possibly be detected by space-based interferometers. Both tensor and scalar perturbations give rise to anisotropy in the temperature of the Cosmic Background Radiation (CBR) on angular scales from arcminutes to 180° (see Fig. 1), and CBR-anisotropy is a promising means of detecting them [4].

Long before inflation the attractiveness of scale-invariant perturbations and a flat Universe had been emphasized [5]; thus tensor perturbations play a crucial role in discriminating between inflation and other "attractive" theories. Further, detection of the tensor perturbations is vital to the reconstruction of the inflationary potential [6]. Denoting the spectral indices of the scalar- and tensor-metric perturbations by $n$ and $n_T$ respectively (scale invariance corresponds to $n - 1 = n_T = 0$), and their contributions to the variance of the quadrupole CBR anisotropy by $S$ and $T$, the value of the inflationary potential and its first two derivatives are given by [7]

$$\begin{aligned}
V(\phi_{50})/m_{\rm Pl}{}^4 &= 1.65(1 - 1.4n_T)\,T, \\
V'(\phi_{50})/m_{\rm Pl}{}^3 &= \pm 8.3\sqrt{\frac{r}{7}}\,T, \\
V''(\phi_{50})/m_{\rm Pl}{}^2 &= 21\left[(n-1) + 0.4r\right]T
\end{aligned} \quad (1)$$

where $r \equiv \frac{T}{S}$. Subscript '50' indicates the value of the inflaton field $\phi$ when present horizon-sized metric fluctuations crossed outside the horizon during inflation; this typically occurs when the scale factor was $e^{50}$ smaller than its value at the end of inflation. In addition, there is a consistency relation: $n_T = -\frac{1}{7}r$, which is an important test of inflation [8].

CBR-anisotropy measurements may offer the best possibility for revealing the presence of tensor perturbations [9]. Because the predictions for the metric perturbations are statistical in nature and the sky is but a finite sample of the Universe, sampling variance, or cosmic variance [10], provides a fundamental limit to the separation of the tensor and scalar contribution to CBR anisotropy. In this *Letter* we show that sampling variance precludes the detection of tensor perturbations if $r \lesssim 0.2$, and further, we show how a simple experiment involving measurements of CBR anisotropy on angular scales of around 3° and 0.5° can achieve this limiting sensitivity.

At present, imprecise knowledge of cosmological parameters, Hubble constant, $H_0 \equiv 100h \, \mathrm{km \, s^{-1} \, Mpc^{-1}}$, baryon fraction, $\Omega_B$, scalar spectral index $n$, and cosmological constant, as well as the ionization history of the Universe limit the sensitivity further, as has been emphasized in Refs. [11]. Further, anisotropy measurements are still dominated by experimental error rather than cosmic variance. We believe that prospects for determining the cosmological parameters by other means is good and experiments are rapidly becoming more accurate, so we take the optimistic view that the limiting factor may soon be sampling variance. For definiteness, we take $h = 0.5$, $\Omega_B = 0.05$, $\Lambda = 0$, $n = 1$, and assume standard ionization history.

We begin by briefly reviewing the statistics of CBR anisotropy in general [12], and that produced by inflation in particular. CBR-temperature fluctuations are expanded in spherical harmonics,

$$\delta T(\theta, \phi) = \sum_{l,m} a_{lm} Y_{lm}(\theta, \phi). \tag{2}$$

Isotropy in the mean guarantees that $\langle a_{lm} a^*_{l'm'} \rangle = C_l \delta_{ll'} \delta_{mm'}$, where brackets indicate average over an ensemble of observers. It is the variance of the multipoles that encodes information about the metric perturbations, and $C_l \equiv \langle |a_{lm}|^2 \rangle$ is called the angular-power spectrum. (The expectation for the square of the quadrupole anisotropy $Q^2 \equiv 5C_2/4\pi$.) Provided that the underlying perturbations are Gaussian (as is almost certainly the case for inflation), all predictions can be derived from the angular-power spectrum. With access to only one sky, for each $l$ we can only measure $2l+1$ independent multipoles. Thus, the estimator for $C_l$, $C_l^{\mathrm{sky}} \equiv \sum_m |a_{lm}|^2/(2l+1)$, will differ from $C_l$ due to finite sampling:

$$\langle (C_l^{\mathrm{sky}} - C_l)^2 \rangle = \frac{2C_l^2}{\text{no. of samples}} = \frac{C_l^2}{l + 1/2}. \tag{3}$$

This ultimate uncertainty in our knowledge of the angular-power spectrum is why there is a limiting sensitivity to $r$.

Most experiments do not directly measure the angular-power spectrum, but instead, the square temperature fluctuation on a given angular scale, whose expectation is

$$\langle \delta T^2 \rangle = \sum_l \frac{2l+1}{4\pi} C_l W_l. \tag{4}$$

The window function $W_l$ depends on the beam size and chopping strategy. Very roughly, an experiment that measures the temperature difference between directions separated by angle $\theta$ with beam size $\sigma \sim \theta$ has a window function that is centered around $l \sim \pi/\theta$, with width of order $l$. If the experiment samples the full sky, then the variance of $\langle \delta T^2 \rangle_{\rm sky}$ is

$$\langle (\langle \delta T^2 \rangle_{\rm sky} - \langle \delta T^2 \rangle)^2 \rangle = 2 \sum_l \frac{2l+1}{(4\pi)^2} C_l^2 W_l^2. \tag{5}$$

The subscript "sky" distinguishes sky average from ensemble average $\langle \delta T^2 \rangle$.

Figure 1 shows the angular-power spectra arising from scale-invariant scalar and tensor perturbations [13]. The expectation for an experiment is determined by the sum of the scalar and tensor contributions, $C_l = C_l^S + C_l^T$. (Likewise, $Q^2 = S + T$, where $S = 5C_2^S/4\pi$ and $T = 5C_2^T/4\pi$.) Differences in the metric between two points on the last-scattering surface (Sachs-Wolfe effect) are the dominant contribution to both angular-power spectra at large angles ($l \lesssim 60$); tensor perturbations decrease in amplitude once their wavelengths become smaller than the Hubble radius which explains why the tensor spectrum drops at small angles ($l \gtrsim 60$) [14]. The scalar angular-power spectrum increases slowly with $l$ due to the pre-recombination oscillations of the baryon-photon fluid, giving rise to the "Doppler peak" at $l \sim 200$ [15].

A simple strategy for separating $C_l^S$ and $C_l^T$ is to measure temperature fluctuations at two angular scales. The window function for one experiment, $W_l^B$, is centered at small angles ($l \sim 200$) where $C_l^T$ is insignificant. The window function for the other experiment, $W_l^A$, is centered at intermediate angles ($l \sim 50$). In essence, experiment B measures the amplitude of the scalar spectrum, and experiment A measures "excess power" at small $l$ due to tensor perturbations. We define a measure of this excess power,

$$Z \equiv \frac{\langle \delta T_A^2 \rangle_{\rm sky}}{\langle \delta T_A^2 \rangle_{r=0}} - 1, \tag{6}$$

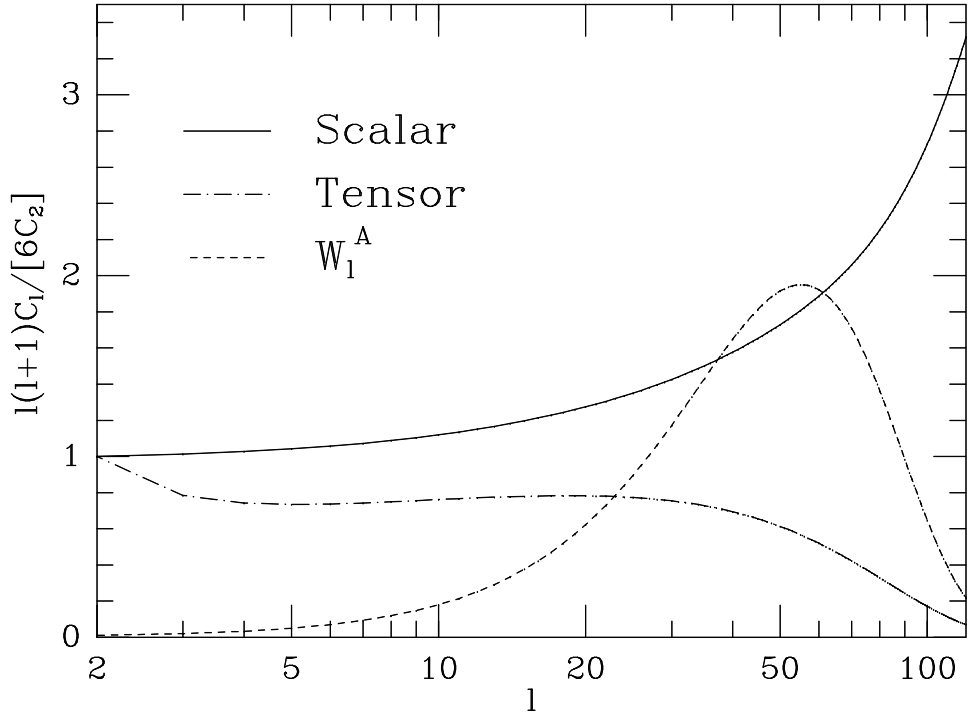

Figure 1: Angular-power spectra arising from scalar and tensor metric perturbations for $n - 1 = n_T = 0$, $h = 0.5$, and $\Omega_B = 0.05$. The dashed curve is the optimal window function, $W_l^A$, described in the text.

where $\langle \delta T_A^2 \rangle_{r=0} = \sum_l \frac{2l+1}{4\pi} C_l^S W_l^A$ is the anisotropy expected in experiment A for $r = 0$ and $\langle \delta T_A^2 \rangle_{\rm sky}$ is that observed. Note that $Z$ is proportional to $r$.

Because of cosmic variance a measurement of $Z \neq 0$ does not necessarily imply $r \neq 0$. Consider the cosmic variance in $Z$:

$$\frac{\Delta Z^2}{\langle Z \rangle^2} \equiv \frac{\langle (Z - \langle Z \rangle)^2 \rangle}{\langle Z \rangle^2} = \frac{2 \sum_l [(2l+1)(C_l^T(r) + C_l^S)^2 (W_l^A)^2]}{(\sum_l [(2l+1) C_l^T(r) W_l^A])^2}. \quad (7)$$

(We have not included the cosmic variance from experiment B because in the case of a full-sky measurement it is negligible.) The ratio $Z/\Delta Z$, which is proportional to $r$, is a measure of signal-to-noise. Choosing a standard form for the window function, $W_l^A = [1 - P_l(\cos \theta)] \exp(-l(l+1)\sigma^2)$, which corresponds to Gaussian beamwidth $\sigma$ and chop angle $\theta$, we maximized $Z/\Delta Z$ (i.e., sensitivity to $r$) by varying $\theta$ and $\sigma$. The optimal window peaks around $l = 50$; see Fig. 1. Its shape can be understood in terms of signal-to-noise weighting: For small $l$ there is large noise (cosmic variance); for large $l$ there is small signal ($C_l^T/C_l^S$ is small).

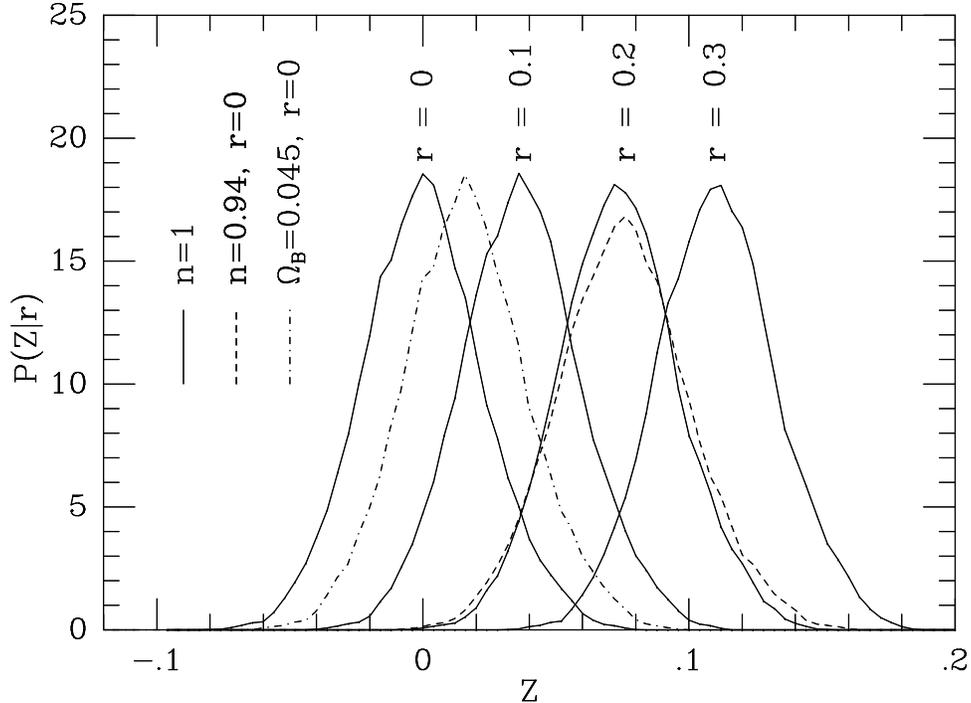

Figure 2: Probability distribution $P(Z|r)$ for $(n, r) = (1, 0)$, $(1, 0.1)$, $(1, 0.2)$ and $(1, 0.3)$ (solid); for $(n, r) = (0.94, 0)$ (broken); and for $(n, r, \Omega_B) = (1, 0, 0.045)$ (dot-dash).

We have numerically calculated the probability distribution $P(Z|r)$ using a Monte-Carlo method; see Fig. 2. For $r \sim 0.3$ it is clearly possible to rule out the hypothesis that $r = 0$, since its probability distribution has little overlap with that for $r = 0.3$.

To address the overlap of distributions quantitatively, statisticians define "size" and "power". The power is the probability, of measuring $Z < Z_1$ given $r = 0$, and the size is the probability of measuring $Z < Z_1$ given $r = r_1$. Figure 3 illustrates the likelihood for ruling out $r = 0$ with 95 per cent confidence, given that the actual value $r \neq 0$. In other words, we have fixed the power to be 0.95, and calculated the size as a function of $r$. As $r$ increases, $P(Z|r)$ overlaps less with the distribution for $r = 0$, and thus the size decreases. For $r \simeq 0.2$ there is a 95 per cent chance that one can exclude the $r = 0$ hypothesis with at least 95 per cent confidence. For the inflationist who feels lucky, for $r = 0.1$ there is a 50 per cent chance of being able to eliminate the $r = 0$ hypothesis with at least 95 per cent confidence.

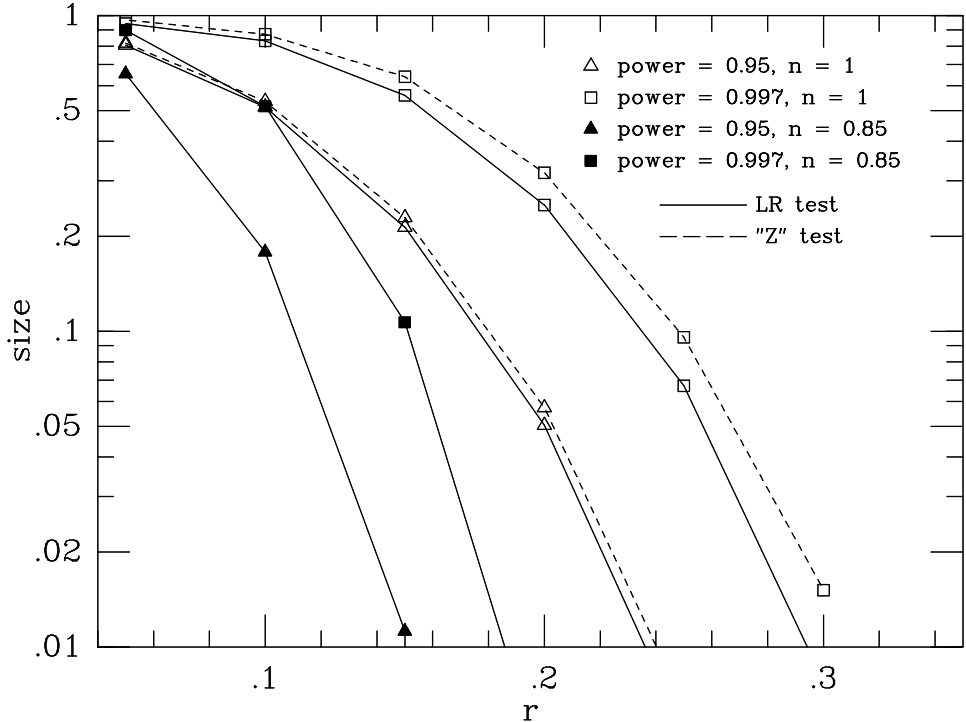

Figure 3: One minus the fraction of sky realizations (= size) that lead to a 95 per cent confidence (power = 0.95) or 99.7 per cent confidence (power = 0.997) exclusion of the $r = 0$ hypothesis as a function of $r$. The broken curves are for the $Z$-test, and the solid curves are for the likelihood-ratio test; $n = 1$ (open symbols) and $n = 0.85$ (filled symbols).

While we have chosen $W_l^A$ to maximize $Z/\Delta Z$, the breadth of this maximum (in window-function space) is large so that even substantial changes in $W_l^A$ do not greatly affect $Z/\Delta Z$. One might ask if a more cleverly defined observable could achieve a greater sensitivity. To the contrary, we now show that the sensitivity cannot be significantly improved.

To address this issue we apply a likelihood-ratio test [16] that uses all
6

possible information, i.e., all the multipoles. Likelihood-ratio tests are designed to discriminate between two hypotheses, $H_0$ and $H_1$. Their virtue is that they are "most powerful;" that is, for fixed size such a test yields maximum power. Let hypothesis $H_0$ be the assertion that $r = 0$, and $H_1$ be the assertion that $r = r_1$.

We define the likelihood ratio,

$$\lambda \equiv \frac{P(\{a_{lm}\}|r = r_1)}{P(\{a_{lm}\}|r = 0)} = e^{-\lambda^*} \prod_{l=2}^{\infty} \left( \frac{C_l^S}{C_l^S + C_l^T(r_1)} \right)^{(2l+1)/2}, \qquad (8)$$

$$\lambda^* = \sum_{l=2}^{\infty} \left[ \frac{\sum_m |a_{lm}|^2}{2} \left( \frac{1}{C_l^S + C_l^T(r_1)} - \frac{1}{C_l^S} \right) \right]. \qquad (9)$$

We then proceed as before: generation of the distributions $P(\lambda|r = 0)$ and $P(\lambda|r = r_1)$, followed by calculation of size for a given power. The results are shown in Fig. 3 and are nearly identical to those of the previous "Z-test."

In principle, all the multipoles relevant for the likelihood-ratio test could be obtained from an all-sky experiment on the 0.5° scale. This does not mean that the virtues of the "Z-test" are solely pedagogical. Its sensitivity is almost the same as the likelihood-ratio test, and full-sky coverage may not be possible, making the extraction of multipoles from any experiment difficult. In the case of partial-sky coverage, the "Z-test" may offer a practical alternative whose limiting sensitivity scales roughly as $[1/f_A + 1/(16f_B)]^{1/2}$ where $f_A$ and $f_B$ are the fractional sky coverages of experiments A and B.

Depending on the actual values of $h$, $\Omega_B$, $\Lambda$, and $n$ the ultimate sensitivity to $r$ varies; it is most sensitive to the shape of the scalar spectrum on the Sachs-Wolfe plateau—and hence $n$. Decreasing $n$ from 1 to 0.85 improves the limiting sensitivity to $r \sim 0.1$ (see Fig. 3); increasing $n$ has the opposite effect. The limiting sensitivity is insensitive to $\Omega_B$ and $h$: for $\Omega_B = 0.02$ it decreases to $r \simeq 0.19$, and for $h = 0.4$ it increases to $r \simeq 0.21$.

What about the effect of imprecise knowledge of cosmological parameters? Figure 2 shows that it is not possible to distinguish $n = 0.94$, $r = 0$ from $n = 1.0$, $r = 0.2$. Suppose $n \simeq 1$, but one could not preclude a value as small as 0.94, the limiting sensitivity increases to $r \sim 0.4$. Inferring extra power on the Sachs-Wolfe plateau (i.e., $r \neq 0$) requires accurate knowledge of the height of the Doppler peak, which depends upon $\Omega_B$ and $h$. Thus, uncertainty in $\Omega_B$ and $h$ can mimic $r \neq 0$. For example, $\Omega_B = 0.045$, $r = 0$ mimics $\Omega_B = 0.05$, $r = 0.04$ (see Fig. 2), and a 10% uncertainty in $\Omega_B$ increases the limiting sensitivity to about 0.24. The effect for $h$ is even greater: $h = 0.55$, $r = 0$ cannot be distinguished from $h = 0.5$, $r = 0.14$.

A logical extension of the two-scale measurement is the extraction of both $r$ and $n_T$ via a three-scale measurement. An independent measure of $n_T$ permits the testing of the consistency relation: $r = -7n_T$. However, even for $r$ as large as 1, so that $n_T = -1/7$, ruling out $n_T = 0$ is nearly impossible. For $r \lesssim 1$, falsification of the consistency relation requires $|n_T| \gg r/7$ [17].

Finally, while we have focussed solely upon anisotropy, polarization is another way of separating tensor and scalar perturbations. For $l \lesssim 10$, linear polarization of CBR anisotropy due to tensor perturbations is about $\sqrt{60r}$ times greater than that due to scalar perturbations [18]. By means of a likelihood-ratio test for polarization and assuming an experiment only limited by cosmic variance, we find a sensitivity of $r \simeq 0.03$. However, the polarization-to-anisotropy ratio is less than 1 per cent for $r = 1$—and closer to 0.01% for $r \simeq 0.03$—and polarization is difficult to measure on large-angular scales. For the foreseeable future it seems unlikely that polarization experiments can achieve higher sensitivity.

In summary, CBR anisotropy appears to be the most promising means of discovering inflation-produced tensor perturbations. We have shown that even with perfect knowledge of the temperature of our CBR sky and the cosmological parameters, the ratio of tensor to scalar perturbations $r$ must be greater than about 0.2 to guarantee a statistically significant detection. The importance of discovering tensor perturbations provides further motivation for the cosmic-variance limited, full-sky CBR anisotropy experiments being planned for satellite platforms.

We thank Scott Dodelson for valuable comments and his scalar angular power spectra. This work was supported in part by the DOE (at Chicago and Fermilab) and by the NASA through grant NAGW-2381 (at Fermilab).

# References

[1] A. H. Guth and S.-Y. Pi, *Phys. Rev. Lett.* **49**, 1110 (1982); S. W. Hawking, *Phys. Lett. B* **115**, 295 (1982); A. A. Starobinskii, *ibid* **117**,

175 (1982); J. M. Bardeen, P. J. Steinhardt, and M. S. Turner, *Phys. Rev. D* **28**, 697 (1983).

[2] V.A. Rubakov, M. Sazhin, and A. Veryaskin, *Phys. Lett. B* **115**, 189 (1982); R. Fabbri and M. Pollock, *ibid* **125**, 445 (1983); A.A. Starobinskii *Sov. Astron. Lett.* **9**, 302 (1983); L. Abbott and M. Wise, *Nucl. Phys. B* **244**, 541 (1984).

[3] N. Kaiser et al., *Mon. Not. R. astr. Soc.* **252**, 1 (1991); M. A. Strauss et al., *Astrophys. J.* **397**, 395 (1992); A. Dekel *Ann. Rev. Astron. Astrophys.*, in press (1994). For a contrary view see, P. Coles and G. Ellis, *Nature* **370**, 609 (1994).

[4] See e.g., R. Crittenden et al., *Phys. Rev. Lett.* **71**, 324 (1993).

[5] R.H. Dicke and P.J.E. Peebles, in *General Relativity: An Einstein Centenary*, eds. S.W. Hawking and W. Israel (Cambridge Univ. Press, Cambridge, 1979), p. 505; E.R. Harrison, *Phys. Rev.* **D1**, 2726 (1970); Ya.B. Zel'dovich, *Mon. Not. R. astr. Soc.* **160**, 1P (1972).

[6] E.J. Copeland, E.W. Kolb, A.R. Liddle, and J.E. Lidsey, *Phys. Rev. Lett.* **71**, 219 (1993); *Phys. Rev. D* **48**, 2529 (1993); M.S. Turner, *ibid*, 3502 (1993).

[7] M.S. Turner, *Phys. Rev. D* **48**, 5539 (1993).

[8] This "consistency relationship," which is accurate to lowest order in $(n-1)$ and $n_T$, holds generally for single-field models; in many models $(n-1) \simeq n_T$ and the stronger relation, $r = -7(n-1)$, also holds; see Refs. [6].

[9] The possibility that the CBR anisotropy discovered by COBE could be due in part to tensor modes was pointed out by L. Krauss and M. White, *Phys. Rev. Lett.* **69**, 869 (1992); R. Davis et al., *ibid*, 1856 (1992); D. Salopek, *ibid*, 3602 (1992); F. Lucchin, S. Mattarese, and S. Mollerach, *Astrophys. J.* **401**, L49 (1992); A. Liddle and D. Lyth, *Phys. Lett. B* **291**, 391 (1992); T. Souradeep and V. Sahni, *Mod. Phys. Lett.* **A7**, 3541 (1992); J.E. Lidsey and P. Coles, *Mon. Not. R. astr. Soc.* **258**, 57P (1992).

[10] See e.g., M. White, L. Krauss, and J. Silk *Astrophys. J.* **418** 535 (1993).

[11] See e.g., J.R. Bond et al., *Phys. Rev. Lett.* **72**, 13 (1994); M. White, L. Krauss, and J. Silk *Astrophys. J.* **418** 535 (1993).

[12] See e.g., J.R. Bond and G. Efstathiou, *Mon. Not. R. astr. Soc.* **226**, 255 (1987).

[13] The scalar angular-power spectrum is from S. Dodelson and J.M. Jubas, *Phys. Rev. Lett.* **70**, 2224 (1993), and the tensor from S. Dodelson, L. Knox and E.W. Kolb, *ibid* **72**, 3443 (1994).

[14] M.S. Turner, M. White and J. Lidsey, *Phys. Rev. D* **48**, 4613 (1993).

[15] W. Hu and N. Sugiyama, astro-ph/9407093 (1994).

[16] See e.g., A.C.S. Readhead et al., *Astrophys. J.* **346**, 566 (1989).

[17] S. Dodelson, L. Knox and E.W. Kolb, *Phys. Rev. Lett.* **72**, 3443 (1994).

[18] R. Crittenden, R. Davis and P. Steinhardt, *Astrophys. J.* **417**, L13 (1993); R.A. Frewin, A.G. Polnarev and P. Coles, *Mon. Not. R. astr. Soc.* **266**, L21 (1994).